\documentclass[letter]{aa}
\usepackage[varg]{txfonts}
\usepackage{graphicx}
\usepackage{amsmath}
\usepackage{longtable}
\usepackage{hyperref}
\hypersetup{colorlinks=true, linkcolor=blue!50!black, citecolor=blue!50!black,
            urlcolor=blue!50!black}

\graphicspath{{Figures/}{Figures/eros_scaled_letter_sc_hyb_px_n20_c15/}}

\newcommand{\msun}{M_{\odot}}
\newcommand{\sigfifty}{\Sigma_{50}}
\newcommand{\rfifty}{r_{50}}
\newcommand{\avir}{\alpha_{\rm vir}}

\newcommand{\sigoned}{\sigma_{\rm 1D}}
\newcommand{\sigbirth}{\sigma_{\rm birth}}
\newcommand{\kms}{km\,s$^{-1}$}

\begin{document}

\title{The dispersal law of young star clusters: free expansion}
\subtitle{A \textit{Gaia/eROSITA} census of Orion and Sco--Cen}
\titlerunning{The dispersal law of young clusters}
\authorrunning{Alves et al.}

\author{Jo\~ao Alves\inst{1}
   \and Stefan Meingast\inst{1}
   \and Alena Rottensteiner\inst{1}
}

\institute{University of Vienna, Department of Astrophysics,
           T\"urkenschanzstra{\ss}e 17, 1180 Vienna, Austria\\
           \email{joao.alves@univie.ac.at}
}

\date{Received ...; accepted ...}

\abstract{%
We measure how young star clusters disperse over time in the two nearest star-forming complexes. Our sample comprises 44 \texttt{SigMA}-\textit{Gaia}-defined clusters in Orion and 31 in Sco--Cen, with isochronal ages of 2--25\,Myr. We extend each cluster's census with \textit{eROSITA}-selected young stars that share its parallax and motion and lie within the distance they could have traveled ballistically since birth. By 10--30\,Myr, about one in three former cluster stars lies outside the cluster's cataloged boundary. Across both complexes and the full age range, we find that (1) the half-member radius increases linearly with age, $\rfifty \simeq \sigbirth t$, (2) the velocity dispersion shows no measurable evolution, and (3) all studied clusters are unbound. These results support free expansion. Young clusters in Orion and Sco-Cen emerge unbound from their birth clouds and coast at their birth velocity dispersion without significant relaxation or a characteristic dispersal timescale. We further show that memberships defined by density contrast preferentially omit the fastest stars, making the same clusters appear to expand more slowly ($\propto t^{0.6}$), cool with age, and approach equilibrium.

}


\maketitle
\raggedbottom

\section{Introduction}
\label{sec:intro}

Most stars form in clusters, yet few survive as bound systems.\footnote{``Cluster'' is used throughout in the statistical sense: an overdensity of stars in position and velocity space, comoving and coeval, with no assumption that it is or ever was gravitationally bound.} In the solar neighborhood, embedded clusters form at more than ten times the rate at which bound open clusters accumulate \citep{lada_embedded_2003}, and only about one in ten cataloged clusters within 250\,pc is compatible with being bound \citep{hunt_improving_2024}. The disappearance of the remainder, known as ``infant mortality'', was long attributed to natal-gas expulsion, which can unbind a cluster within a few Myr \citep{hills_effect_1980, lada_formation_1984, kroupa_formation_2001, goodwin_gas_2006}. \citet{krumholz_star_2019} framed this rapid dispersal as the first phase of a life cycle that ends, for the bound minority, in Gyr-scale tidal disruption.

However, the time evolution of dispersal in an unbound young cluster has not yet been measured directly. \textit{Gaia} observations show that most young clusters are expanding, but these studies generally measure a cluster's expansion rate at a single epoch \citep{kuhn_kinematics_2019, della_croce_young_2024, wright_ob_2023}. A kinematic age is then estimated by dividing the cluster's size by its expansion rate, effectively assuming ballistic motion rather than testing it \citep{luhman_census_2023, armstrong_expansion_2024, armstrong_expansion_2026}. N-body models and the weakening of expansion signatures with age support a scenario in which clusters revirialize and their expansion ends within 10--20\,Myr \citep{pfalzner_observations_2019, della_croce_young_2024, pang_different_2020, calovic_longterm_2025}. On the scale of entire OB associations, observations likewise show no coherent expansion from a common compact configuration \citep{wright_kinematics_2018, ward_not_2020}. Thus, previous studies have characterized expansion rates or identified possible dispersal epochs, but none have measured how the size of a young cluster population changes with age.

A key obstacle is the way cluster membership is defined. In a modern catalog, a cluster consists of the stars that a clustering algorithm recovers as an overdensity above the field. Its membership therefore ends where the density contrast disappears. 
Quantities derived from the cataloged membership consequently describe the retained population rather than the population at birth. Open clusters are surrounded by precisely such stars, forming unbound coronae that contain much of their mass \citep{meingast_extended_2021, tarricq_structural_2022}. A corresponding population has been detected in X-rays in Sco--Cen whose origin remains uncertain \citep{schmitt_xraying_2022}. N-body models reinforce this picture and argue that unbound former members preserve essential information about a young cluster's history, whereas cuts in distance or velocity can produce a misleading reconstruction \citep{arunima_unbound_2023}.

\section{Towards a complete census}
\label{sec:data}

\begin{figure*}[t]
\centering
\includegraphics[width=0.98\textwidth]{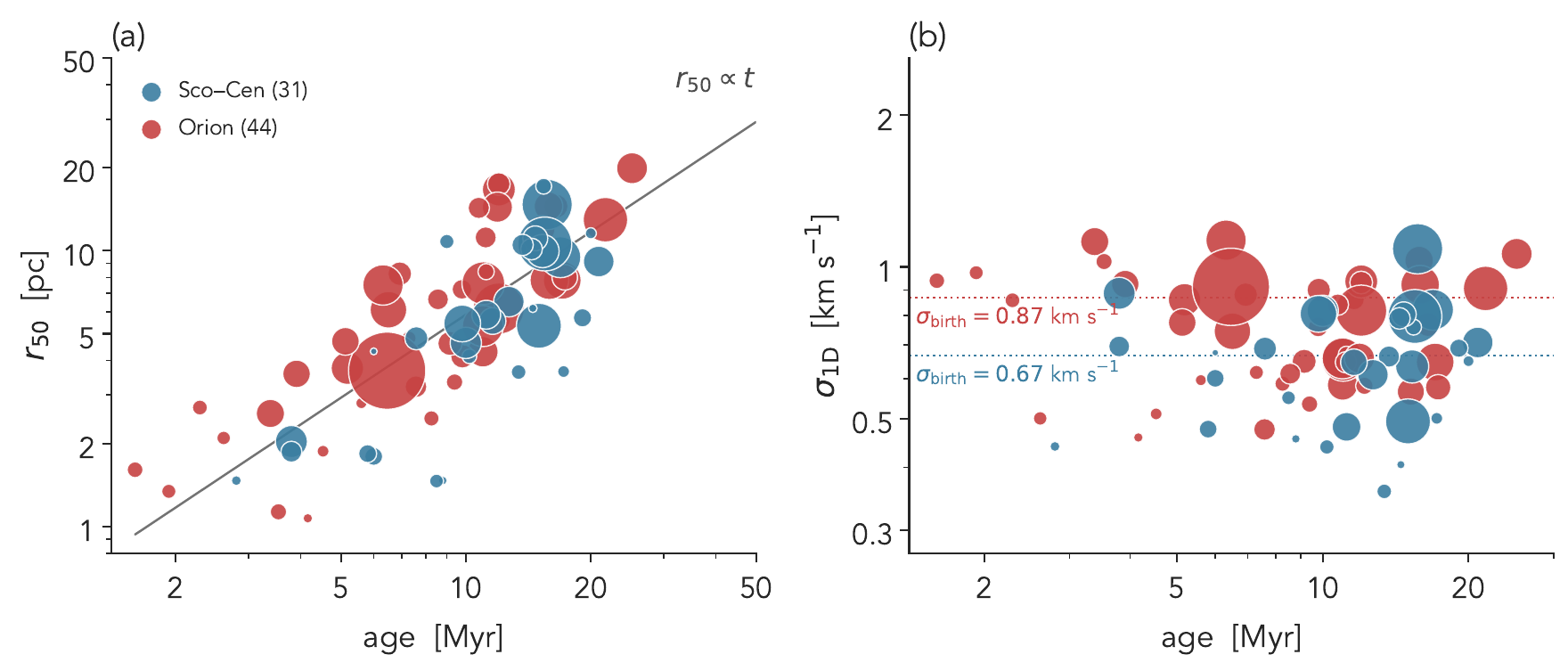}
\caption{(a) Half-member radius of the extended census against isochronal age
for the 75 clusters, with the marker area proportional to $N$. The solid line
is $\rfifty \propto t$ through the joint geometric mean of the ages and radii. 
(b) Tangential dispersion against the same ages. The dotted lines are the two
complexes' birth dispersions $\sigbirth$ (Sect.~\ref{sec:law}).}
\label{fig:law}
\end{figure*}

We adopted two \texttt{SigMA} clustering solutions derived from \textit{Gaia} DR3 astrometry \citep{vallenari_gaia_2023}, using ages and 1$\sigma$ uncertainties that were fitted consistently to PARSEC isochrones. For Sco--Cen we relied on the star-by-star catalog from \citet{grossschedl_evolution_2026}, which merges 34 of the 37 clusters in \citet{ratzenbock_star_2023} with the two TW~Hya clusters identified by \citet{miret-roig_tw_2025}. For Orion, we used the sample of 47 clusters and 11\,996 strict members from \citet{rottensteiner_orion_2026}. Requiring at least 20 strict members excludes the lowest-quality clusters, whose broad search regions would otherwise sweep up nearby field populations, resulting in 44 Orion clusters (1.6--25\,Myr) and 31 Sco--Cen clusters (2.8--20.9\,Myr, Appendix~\ref{app:member}).

We draw candidates from the eRASS:3 X-ray catalog \citep{predehl_erosita_2021, merloni_srg_2024} cross-matched with \textit{Gaia} DR3 \citep{ramos-ceja_srg_2026} in order to extend the existing census. Young stars in each complex are then selected using one of two approaches: when the pre-main sequence is distinct from the main sequence, we keep stars lying above the 40-Myr isochrone. When the two sequences overlap, we instead select based on X-ray activity (Appendix~\ref{app:member}). A star is linked to a cluster if three requirements are met: (1) its parallax agrees with that of the cluster, (2) its tangential velocity matches the cluster's within 3\,\kms, and (3) its position falls within the ballistic reach of the cluster. The ballistic reach in (3) is defined as the birth radius plus three times the distance traveled by a star moving at the birth velocity dispersion over the cluster age.  
Because \textit{eROSITA} does not detect every young star, each attached star enters
every statistic with a weight: the cluster's inverse X-ray detection
fraction, measured on its own members, times a field purity measured in
parallax shells in front of and behind the complex
(Appendix~\ref{app:member}).

For each cluster, we use the \texttt{SigMA} + \textit{eROSITA} extended sample to measure
\begin{equation}
  \rfifty = d_{\rm med}\,\theta_{50},\qquad
  \avir = \frac{5\,\sigoned^{2}\,\rfifty}{G\,N\,\langle m\rangle},
  \label{eq:obs}
\end{equation}
where $\theta_{50}$ is the weighted median angular separation from the cluster center and $d_{\rm med}$ is the weighted median distance. Their product gives the projected half-member radius $\rfifty$ in parsecs. $N$ is the weighted number of stars, and $\langle m\rangle=0.6\,\msun$ is the mean mass of a \citet{kroupa_variation_2001} IMF. We calculate $\sigoned$ from the two plane-of-the-sky LSR velocity components after correcting for measurement errors and the cluster's perspective field (Appendix~\ref{app:sigma}). We exclude radial velocities because binary orbital motion can broaden single-epoch measurements beyond the intrinsic dispersion. We fit all slopes by unweighted least squares in log--log space, using one point per cluster. Uncertainties are reported as $x^{+a}_{-b}$, with bounds given by the 16th and 84th percentiles of bootstrap resamples of the clusters.

\section{Results}
\label{sec:results}

\subsection{Main result: $\rfifty \simeq \sigbirth t$}
\label{sec:law}

Figure~\ref{fig:law}a plots the half-member radius from the extended census as a function of age. In both complexes it increases roughly linearly with age: $\rfifty \propto t^{p}$, with $p = 0.97^{+0.10}_{-0.09}$ for Orion and $1.08^{+0.15}_{-0.12}$ for Sco--Cen. The radius also depends on total number of cluster members (richness) in addition to age, since at a given age richer clusters are larger. We therefore fit $\log\rfifty$ simultaneously as a function of $\log t$ and $\log N$. At fixed richness, the inferred growth exponent is $0.88^{+0.10}_{-0.10}$ in Orion and $1.00^{+0.16}_{-0.13}$ in Sco--Cen, and the richness dependence is similarly weak in both complexes ($b = 0.13$--0.14, compared to 0.32--0.37 when using only the \texttt{SigMA}-assigned members, Appendix~\ref{app:member}).
The proportionality constant is the birth dispersion $\sigbirth$. We measure it on the clusters younger than 5\,Myr, the population closest to birth, as the median $\sigoned$ of their assigned members: 0.87\,\kms\ over nine clusters in Orion and 0.67\,\kms\ over three in Sco--Cen (Appendix~\ref{app:member}).

\subsection{Velocity dispersion does not change}
\label{sec:sigma}

Figure~\ref{fig:law}b shows the tangential dispersion of the same clusters. It does not change with age: $d\log\sigoned/d\log t = -0.03^{+0.07}_{-0.06}$ in Orion and $+0.07^{+0.09}_{-0.09}$ in Sco--Cen.  On the assigned members alone the same clusters give $-0.24$ and $-0.26$ (Sect.~\ref{sec:discussion}). Together with Sect.~\ref{sec:law}, these results imply free expansion.

\subsection{Nothing is bound at the measured scale}
\label{sec:bound}

\begin{figure}
\centering
\includegraphics[width=\columnwidth]{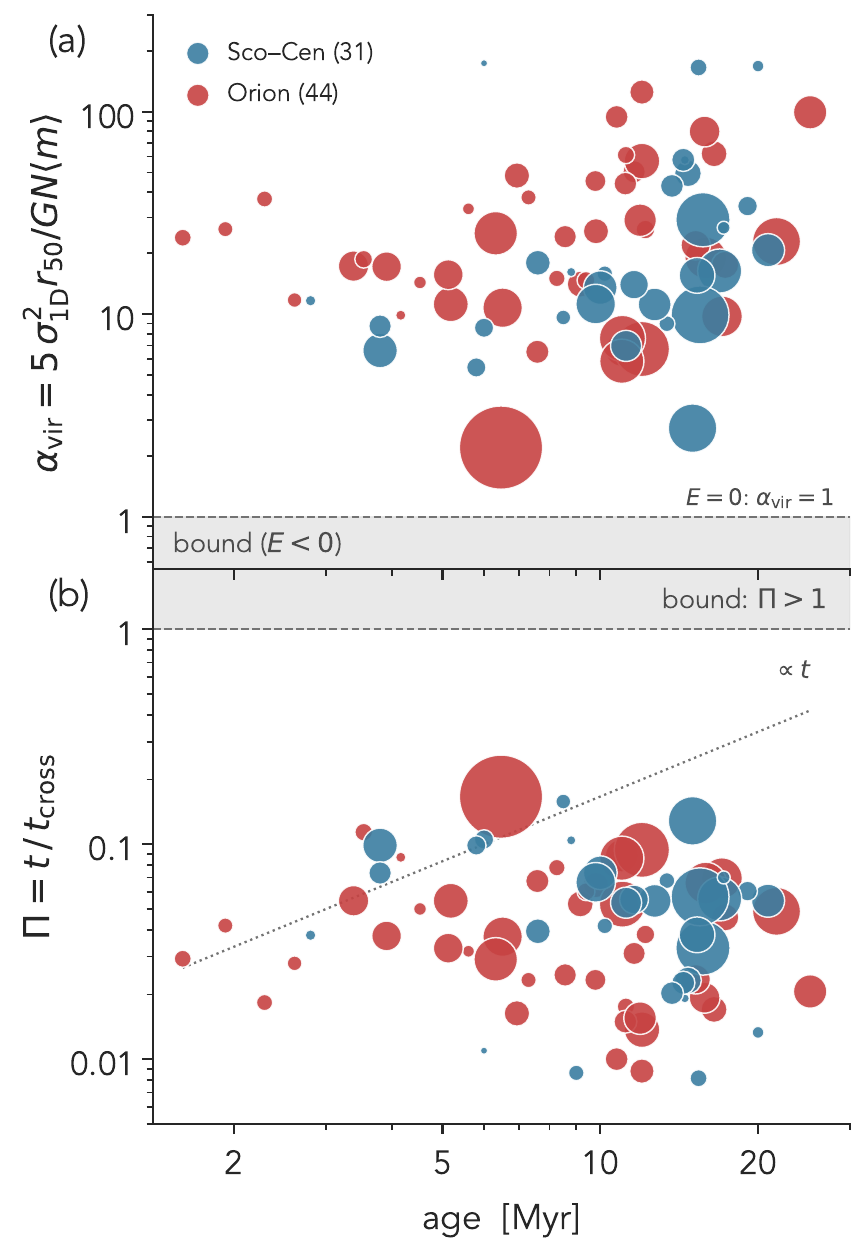}
\caption{(a) Virial ratio of the extended census against age. The dashed line
at $\avir = 1$ is the bound--unbound boundary to within a few percent
(Sect.~\ref{sec:bound}). (b) Age in crossing times. The dotted line is
the bound expectation $\Pi \propto t$ (Sect.~\ref{sec:bound}).
Marker area is proportional to $N$, as in Fig.~\ref{fig:law}.}
\label{fig:bound}
\end{figure}

Figure~\ref{fig:bound}a presents the virial ratio. Since $\rfifty$ is a projected half-member radius rather than a true outer radius, the condition of zero total energy corresponds to $\avir = 0.98$--1.22 depending on the assumed density profile, and the dashed line at $\avir = 1$ therefore indicates the bound--unbound division to within a few percent. All clusters fall above this line. The median value is 22 in Orion and 16 in Sco--Cen, with a minimum of 2.2 for the ISF. The $N$ values count \textit{Gaia} detected sources, and thus omit unresolved companions, embedded members, and gas. Consequently, $\avir$ should be regarded as an upper bound by a factor that we cannot constrain to better than roughly two. For the ISF this factor is comparable to its own offset above the dividing line, and its central region, the ONC, is separately known to be sufficiently dense to be gravitationally bound \citep{hillenbrand_preliminary_1998}. The census characterizes the larger-scale structures and does not rule out bound cores embedded within them.

Figure~\ref{fig:bound}b reaches the same conclusion without the velocity dispersion. The age in crossing times, $\Pi = t/t_{\rm cross}$ with $t_{\rm cross} = 10\,(r_h^{3}/GM)^{1/2}$ and $r_h = 1.305\,\rfifty$ \citep{gieles_distinction_2011}, has a median $\Pi$ of 0.037 in Orion and 0.055 in Sco--Cen, and no cluster exceeds 0.17. The evolution of $\Pi$ is a stronger diagnostic than its absolute value. For a bound, non-expanding cluster, the crossing time remains fixed and $\Pi\propto t$. Free expansion instead predicts $\Pi\propto t^{-1/2}$ at fixed $N$. 
Including the observed richness trend, $d\log N/d\log t=+0.6$, changes this prediction to $\Pi\propto t^{-0.2}$ (Appendix~\ref{app:limit}). The measured slopes are $-0.14^{+0.12}_{-0.16}$ in Orion and $-0.34^{+0.23}_{-0.27}$ in Sco--Cen. Thus, $\Pi$ decreases rather than increasing with age, and no cluster in the census shows evidence of becoming bound.

\section{Discussion}
\label{sec:discussion}

The main result of this work is that most young clusters disperse through free expansion. Across the extended census, the half-member radius follows $\rfifty \simeq \sigbirth t$, the velocity dispersion remains constant with age, and every cluster is unbound. These results hold from 2 to 25\,Myr in Orion and Sco--Cen, two complexes with different cluster richness, distances, and histories. The systems studied here are not a rare evolutionary class. About nine in ten cataloged clusters within 250\,pc are unbound \citep{hunt_improving_2024}, while the bound minority is not represented in our census. Free expansion therefore appears to be the dominant mode of dispersal for nearby young clusters.
For this majority, the expansion detected in single-epoch measurements \citep{kuhn_kinematics_2019, della_croce_young_2024, wright_ob_2023} is not a short-lived phase. It neither slows through relaxation nor ends after 10 to 20\,Myr, but continues across the full age range of the census. A freely expanding population also has no characteristic dispersal timescale. An exponential fit to its declining density recovers not an intrinsic lifetime but an e-folding time close to the mean age of the selected sample. The few-Myr infant-mortality timescale found in the literature is therefore the same relation evaluated for a sample a few Myr old (Appendix~\ref{app:limit}). Rather than a lifetime, an unbound young cluster is characterized by three quantities: its birth radius, its birth velocity dispersion, and the density contrast below which a census no longer identifies it as a cluster.

The observed relations also constrain when and how the expansion began. First, dispersal began at birth. Every cluster has $\Pi \leq 0.2$, so there has been too little time for collisional processes such as relaxation, evaporation, or dynamical ejections to shape its evolution \citep{moeckel_rapid_2012}. Moreover, clusters as young as 1.6 to 2\,Myr already lie on the relation $\rfifty = \sigbirth t$ through the origin. Any bound phase must therefore have lasted less than about 1\,Myr.
Second, the mechanism that launched the expansion acted only once. A constant velocity dispersion, a radius proportional to age, and a decreasing $\Pi$ are the signatures of ballistic coasting. Together, they exclude slow gas loss followed by revirialization \citep{pang_different_2020, calovic_longterm_2025, della_croce_young_2024} and remove the need for accelerated expansion \citep{zamora-aviles_flipping_2019}.
Third, the launch speed is consistent with the gravitational potential of the natal clump. A velocity dispersion of $0.87$\,\kms\ at a birth radius of $0.84^{+0.18}_{-0.21}$\,pc   (the intercept of $\rfifty^{2} = r_{0}^{2} + (vt)^{2}$, the quadrature sum of the birth spread and the distance coasted at speed $v$, fitted over the Orion clusters)  corresponds to a virial mass of $5\sigbirth^{2}r_{0}/G \simeq 740\,\msun$ and a mean density within $r_{0}$ of $n_{\rm H_{2}} \simeq 4\times10^{3}$\,cm$^{-3}$. These are characteristic values for a dense cluster-forming clump. By comparison, the median cluster in the extended census contains about 160\,$\msun$ in stars. The implied effective star formation efficiency \citep{goodwin_gas_2006} is then about 0.2 in Orion and 0.3 in Sco--Cen, and not distinguishable within the errors. These estimates are lower limits because $N$ counts \textit{Gaia} systems rather than individual stars. Stars moving at the virial speed of a clump while contributing only a minority of its mass are in the classical configuration for unbinding through rapid gas expulsion \citep{hills_effect_1980, lada_formation_1984, goodwin_gas_2006}. The present census cannot determine whether the stars were briefly bound before being released or were never bound and inherited the motions of an already dispersing clump. 

The contrast with previous results arises from how cluster membership is defined. When the same 75 clusters are analyzed using only their assigned members, the half-member radius scales as $\rfifty \propto t^{0.63}$ in Orion and $\rfifty \propto t^{0.92}$ in Sco--Cen. At fixed richness, their velocity dispersions decline as $t^{-0.24}$ and $t^{-0.26}$, respectively. At the oldest ages, the clusters also appear to approach equilibrium. These trends reproduce the interpretation of member-based expansion studies \citep{della_croce_young_2024, pfalzner_observations_2019} and help explain why N-body models initialized from cataloged memberships retain bound cores \citep{sanchez-sanjuan_dynamical_2026}.
Because the fastest stars leave first, the cataloged membership becomes increasingly biased toward the slower stars that remain. Quantities derived from that membership, including kinematic ages \citep{luhman_census_2023, armstrong_expansion_2024, armstrong_expansion_2026}, therefore describe the retained population rather than the population at birth (Appendix~\ref{app:member}).

Finally, free expansion at the cluster scale does not require coherent expansion at the scale of an entire OB association. The evidence that associations did not originate in a single compact configuration therefore remains unchanged \citep{wright_kinematics_2018, ward_not_2020}. An association need not expand from a common center because its constituent clusters each expand from their own.

\section{Conclusions}
\label{sec:conclusions}

\begin{itemize}

\item Departed members from the cataloged clusters account for about one third of the young population. When these stars are included, young clusters follow the free-expansion relation $\rfifty \simeq \sigbirth t$. The growth exponents are $0.97^{+0.10}_{-0.09}$ in Orion and $1.08^{+0.15}_{-0.12}$ in Sco--Cen, while the velocity dispersion remains constant with age in both complexes.

\item No cluster is bound or evolving toward a bound state. Every cluster lies above the zero-energy line, and the age in crossing times, $\Pi \leq 0.2$, does not increase in proportion to age as it would for a bound cluster.

\item The expansion was launched in a single rapid event at birth and was completed within one crossing time. Its velocity is consistent with the virial speed of the natal dense clump. The observations exclude slow gas loss, relaxation-driven dispersal, and accelerated expansion. They do not distinguish between clusters that were briefly bound before dispersing and clusters that were never bound. Measurements of the embedded phase are needed to distinguish between these possibilities.

\item Memberships defined by density contrast preferentially retain the slower stars. When only cataloged members are considered, the same clusters grow as $t^{0.6}$, cool with age, and appear to approach equilibrium. Expansion ages, virial analyses, and N-body predictions based on these memberships inherit the same bias.

\end{itemize}

For most stars, the terms ``embedded cluster'' and ``association'' describe the same object at different ages rather than distinct kinds of object.

\begin{acknowledgements}
Co-funded by the European Union (ERC, ISM-FLOW, 101055318). This work has made use of data
from the European Space Agency (ESA) mission \textit{Gaia}
(\url{https://www.cosmos.esa.int/gaia}), processed by the \textit{Gaia}
Data Processing and Analysis Consortium (DPAC,
\url{https://www.cosmos.esa.int/web/gaia/dpac/consortium}).
This work is based on data from eROSITA, the soft X-ray instrument aboard
SRG, a joint Russian-German science mission supported by the Russian Space
Agency (Roskosmos), in the interests of the Russian Academy of Sciences
represented by its Space Research Institute (IKI), and the Deutsches Zentrum
f\"ur Luft- und Raumfahrt (DLR). The SRG spacecraft was built by Lavochkin
Association (NPOL) and its subcontractors, and is operated by NPOL with
support from the Max Planck Institute for Extraterrestrial Physics (MPE). The
development and construction of the eROSITA X-ray instrument was led by MPE,
with contributions from the Dr.\ Karl Remeis Observatory Bamberg \& ECAP (FAU
Erlangen-Nuernberg), the University of Hamburg Observatory, the Leibniz
Institute for Astrophysics Potsdam (AIP), and the Institute for Astronomy and
Astrophysics of the University of T\"ubingen, with the support of DLR and the
Max Planck Society. The Argelander Institute for Astronomy of the University
of Bonn and the Ludwig Maximilians Universit\"at Munich also participated in
the science preparation for eROSITA.
\end{acknowledgements}

\bibliographystyle{aa}
\bibliography{decay_letter}

\begin{appendix}

\section{The velocity dispersion and the virial ratio}
\label{app:sigma}

The dispersion of Eq.~(\ref{eq:obs}) is the root of the mean error-corrected
variance of the two plane-of-the-sky LSR velocity components $v_1$ and $v_2$,
\begin{equation}
\sigoned^{2} = \tfrac{1}{2}\sum_{i\,=\,1,2}\left(s_i^{2} - \epsilon_i^{2}\right).
\label{eq:sig}
\end{equation}
The two terms of Eq.~(\ref{eq:sig}) separate the cluster from the instrument.
Measurement errors scatter each star's velocity about its true value, so the
scatter we observe is the cluster's own dispersion and the errors added in
quadrature. Subtracting $\epsilon_i^2$ leaves the dispersion. The observed
scatter is $s_i = 1.4826\,\mathrm{MAD}(v_i)$, the median absolute deviation of
component $i$ about the cluster's median velocity, rescaled by 1.4826 to the
scale of a standard deviation for a Gaussian. The MAD ignores the tails of the
distribution, so a few binaries or interlopers cannot inflate it. The error
$\epsilon_i$ is the median per-star measurement error of the same component.
On the extended census each attached
star enters the medians and the MADs at the weight defined in
Sect.~\ref{sec:data}, and Eq.~(\ref{eq:sig}) returns the member-only
dispersion when every weight is one. For an isotropic velocity field the
three-dimensional dispersion is $\sqrt{3}\,\sigoned$.

Regarding perspective correction, a cluster whose members share one space
velocity does not show one common proper motion. Each member lies along a
slightly different line of sight, so the shared velocity projects onto the sky
by a different amount for each. The result is an apparent tangential field
$-v_r\theta$ directed at the cluster center, where $\theta$ is the angular
offset of a member and $v_r$ is the systemic radial velocity. This is the
moving-cluster effect in its linear form \citep{de_bruijne_refurbished_1999},
and the separation of a real expansion from the perspective field of an approaching group was the central difficulty of the first kinematic studies of OB associations
\citep{blaauw_o_1964, brown_kinematic_1997}. The field is ordered rather than
random, the same function of position for every member, so we subtract it star
by star rather than in quadrature. Both catalogs give velocities in the LSR
frame, so the relevant $v_r$ is the LSR value: along Orion's line of sight the
solar motion is $-16.7$\,\kms, and $v_r$ is 4.7\,\kms\ rather than the
heliocentric 21.7. In Orion the term is 0.06 of $\sigoned$, and removing it
changes nothing. In Sco--Cen it is 0.51 of $\sigoned$ and exceeds $\sigoned$
itself in 10 of 35 clusters, because the median Sco--Cen cluster subtends
1.9$^\circ$ within its half-member radius against 0.6$^\circ$ in Orion. 

\section{The extended membership}
\label{app:member}

\begin{figure*}[t]
\centering
\includegraphics[width=0.8\textwidth]{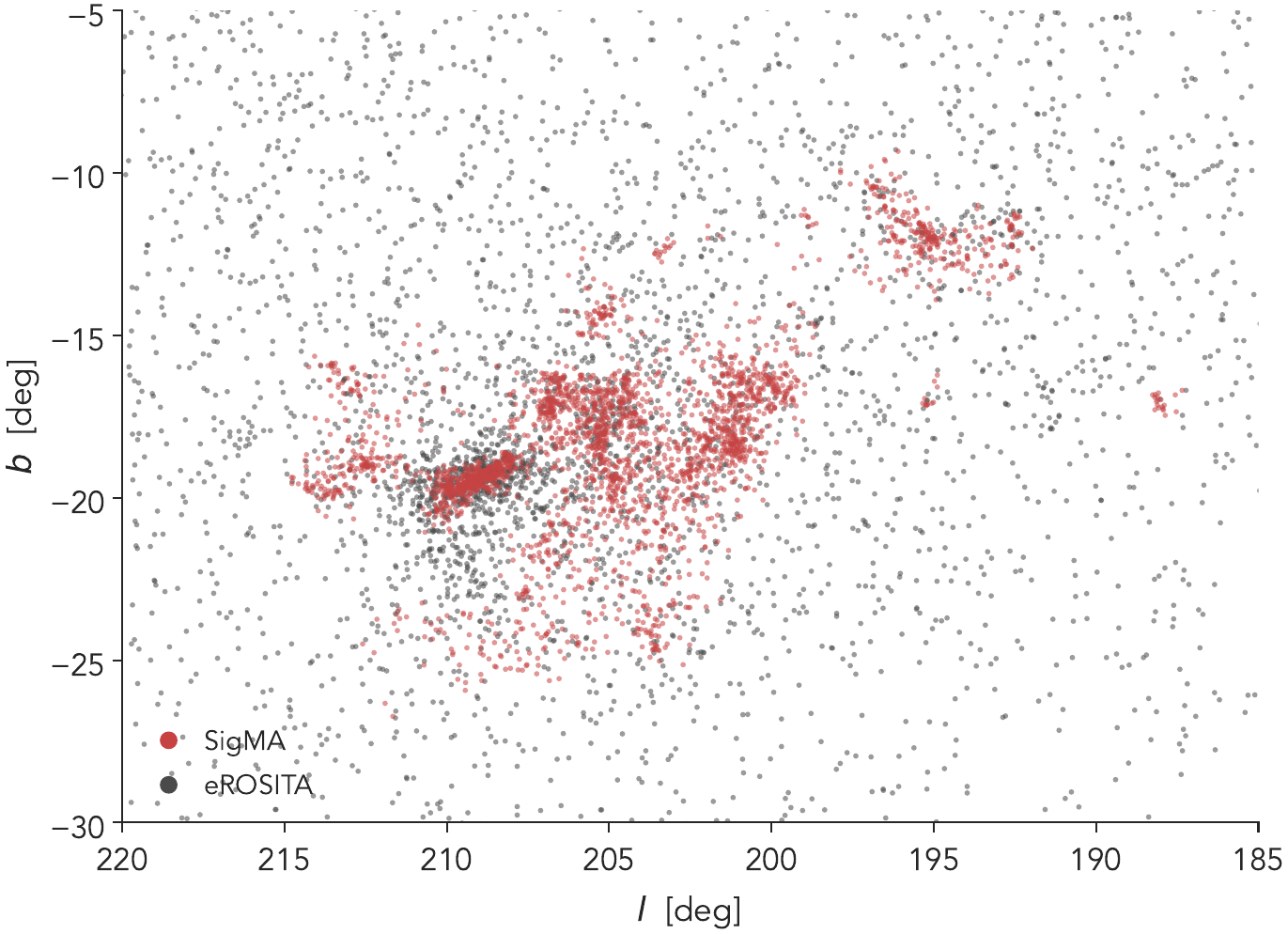}
\caption{Sky positions of the X-ray-detected assigned members (red) and of
all unlisted X-ray-young stars at Orion's parallax (dark gray). The cluster of \textit{eROSITA} sources at about GLON $210^\circ$ and GLAT $-20^\circ$ with few \texttt{SigMA} sources corresponds to the embedded population inside Orion A, unreachable by \textit{Gaia}.}
\label{fig:sky}
\end{figure*}

\begin{figure*}[t]
\centering
\includegraphics[width=0.98\textwidth]{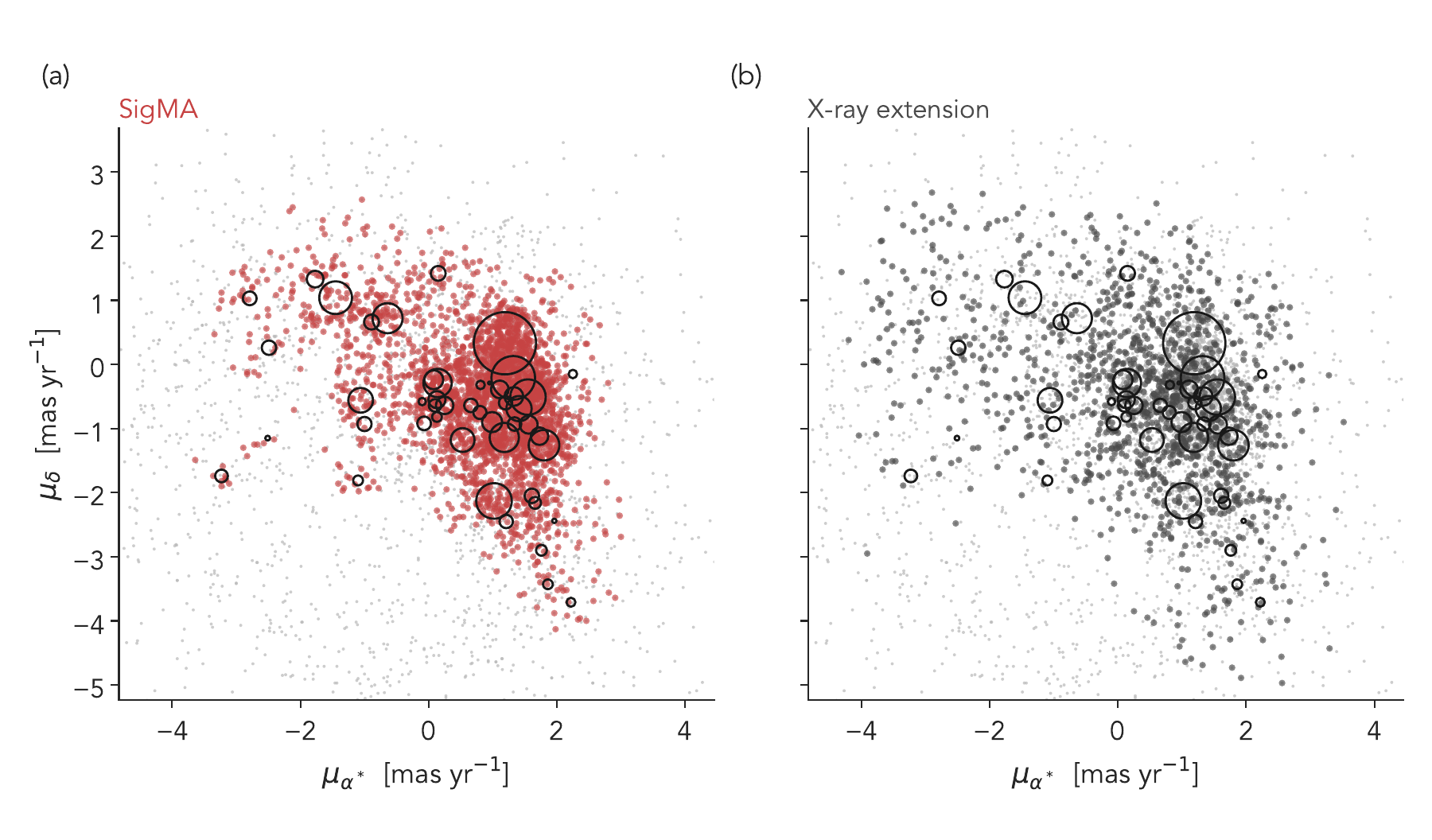}
\caption{Proper motions of the X-ray-detected assigned members (a) and of
the attached departed stars, the X-ray extension of the membership (b). The
light gray background in both panels is the full X-ray-young selection at
Orion's parallax. The open black circles, with area proportional to the
cluster's number of members, mark the cluster mean proper motions.}
\label{fig:pm}
\end{figure*}

\begin{figure*}[t]
\centering
\includegraphics[width=0.98\textwidth]{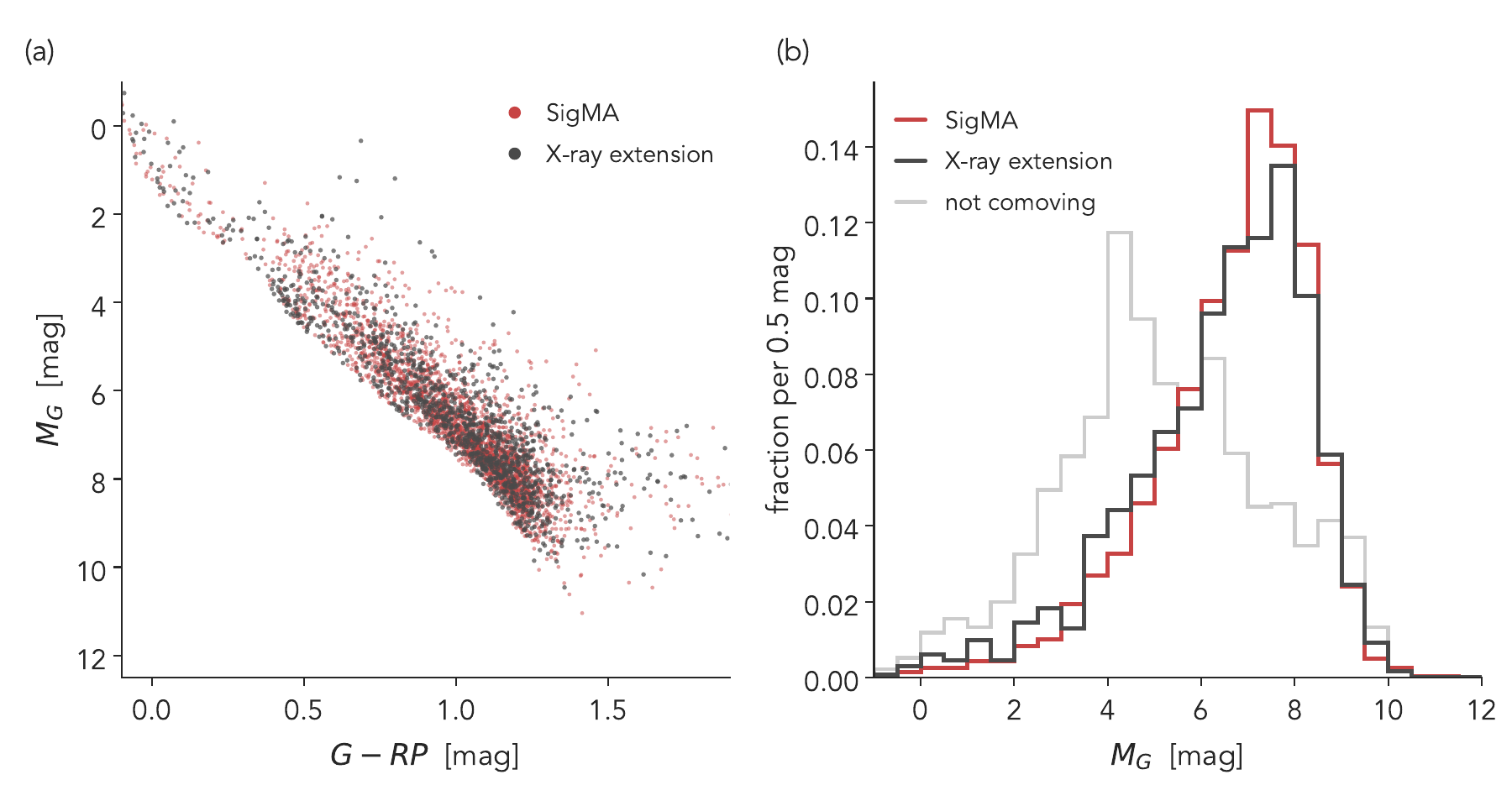}
\caption{(a) Color--magnitude diagram of the X-ray-detected assigned members
(red) and the attached departed stars (dark gray). (b) Luminosity functions of the
assigned, the departed and the not-comoving stars as fractions per 0.5-mag
bin.}
\label{fig:cmdlf}
\end{figure*}

\subsection{The selection} Young, low-mass stars emit strong X-rays because they are still rapid rotators, and rapid rotation produces a corona much brighter than that of an older star. This elevated emission persists for tens of Myr, meaning an X-ray detection can identify a star as young well after its birth cloud has dispersed. We use the eRASS:3 X-ray source catalog with \textit{Gaia} DR3 counterparts \citep{ramos-ceja_srg_2026}, restricted to the parallax range of each complex.

Youth is then verified in one of two ways, depending on the star's color, since no single diagnostic applies across all masses. (1) For $0.8 < BP-RP < 4$, we use the color--magnitude diagram. Because a pre-main-sequence star is still contracting, it is more luminous than an older star at the same color, so we require it to fall above the 40-Myr PARSEC isochrone. Over this color interval, the young and old loci are separated by more than the photometric scatter. (2) For stars bluer than $BP-RP = 0.8$, this criterion breaks down: near 1\,$\msun$ the pre-main sequence has essentially merged with the main sequence and the isochrones no longer diverge. In that regime, X-ray luminosity becomes the discriminator. Young stars have saturated coronae whereas older stars' coronae have weakened, producing differences of orders of magnitude, so we require $L_X/L_{\rm bol} > 10^{-5}$ for $0.3 \leq BP-RP < 0.8$. For $BP-RP < 0.3$, the star is type A or earlier and lacks an intrinsic corona, so an X-ray detection implies an unresolved low-mass companion. We therefore impose only $L_X > 3\times10^{29}$\,erg\,s$^{-1}$. In both blue-color cases, the star must also fall within a magnitude band around the 40-Myr isochrone, extending from 2\,mag above to 1.5\,mag below. This band excludes giants and active subgiant binaries above it, as well as X-ray-bright white dwarfs and hot subdwarfs that lie several magnitudes below.

The boundary isochrone age is set by two competing constraints. It has to be older than the oldest clusters in the census (25\,Myr) by a sufficient margin that their members still lie above it once photometric scatter, unresolved binaries, and model imperfections are taken into account. At the same time, it should be no older than necessary, because each additional Myr in the boundary only lets in more older field stars while adding no new members, given that the census contains nothing older than 25\,Myr. The boundary is thus chosen as the youngest isochrone that the oldest members still clear. At 30\,Myr they do not: near $BP-RP = 2$, a 25-Myr star lies within about a tenth of a magnitude of that curve, well within the members' spread about their own isochrone, so the cut would run through the 10--30\,Myr clusters where the fraction lost is greatest. At 40\,Myr, the same star sits a few tenths of a magnitude above the line across the full color range, and we adopt that as the boundary. Between the two possible mistakes, admitting field stars is less costly: their contamination can be estimated from the parallax shells used for the weighting below and removed, whereas a true member excluded by the cut cannot be regained.

Figures~\ref{fig:sky} and \ref{fig:pm} show the selection on the sky and in
proper motion. Every star drawn there is an eRASS:3 detection, so the members
shown are the 2\,790 X-ray-detected stars among the 11\,996 strict members,
the $A$ of the weighting below (Appendix~\ref{app:weight}). Figure~\ref{fig:cmdlf} compares the assigned
and the attached stars in the color--magnitude diagram. Their luminosity
functions agree in shape (Fig.~\ref{fig:cmdlf}b), so the two are drawn from
one population, and the weighting below is justified.

\subsection{The attachment} A former member keeps three properties of its
cluster. It still sits at the cluster's distance, it still moves with the
cluster to within the speed it left with, and it cannot have traveled
farther than that speed times the cluster's age. The attachment tests all
three.

(1) Distance. The star's parallax must agree with the cluster's to within
three times the quadrature sum of the star's parallax error and the
cluster's own parallax spread, the robust scatter of its strict members.
The test removes stars that share the cluster's sightline at another
distance.

(2) Velocity. The star's tangential velocity, evaluated at the cluster's
distance, must match the cluster's to within 3\,\kms. The window is wide
for a member: the members' velocity dispersion is about 0.8\,\kms, so 3\,\kms\ is
close to four times that and includes roughly 98\% of them. It is tight for an
escaped star, since the stars that depart tend to be the high-velocity ones. Their
speed distribution is determined from the X-ray-young stars around the
clusters, in each speed bin, using two independent indicators of past membership
that yield the same result. A former member tends to move away from its cluster, while a field star has no preferred direction. Within a given speed bin, the outward-going fraction therefore quantifies the fraction of former members. Likewise, a former member follows the members' luminosity function, whereas a field star follows the field's, inferred from stars at 15--30\,\kms. Therefore, fitting each speed bin with a combination of these two distributions yields the same fraction.
The departed stars have a
median speed of 2.0\,\kms\ and a tail to 10\,\kms. The 3\,\kms\ window
therefore contains 67\% of them at a purity of 95\%, and a 5\,\kms\ window
would contain 85\% at 88\%. We keep 3\,\kms\ as the purer choice and report
the 5\,\kms\ result in the robustness section below.

(3) Reach. The star must lie within the distance a member could have
traveled: a starting radius of 1.5\,pc plus the drift of a star moving at
three times the birth dispersion, $3 \times 0.8\,{\rm km\,s^{-1}} \times t$,
for the cluster's age $t$. The factor three is a compromise. It admits the
departed stars that left near the birth dispersion and truncates the fast
tail, and the robustness section below reports what happens at two and at
four. Two simpler choices fail. A fixed physical radius lets a 2-Myr
cluster attach distributed stars 25\,pc away, which no member could have
reached without moving at 12\,\kms. A radius set by the cluster's cataloged
extent grows with the very quantity the attachment is meant to correct. The
reach is therefore set by the physics and by nothing in the catalog. On the
sky it is capped at $15\degr$, beyond which the tangential velocity is no
longer a small-angle projection and its error exceeds 1\,\kms.

A star that passes all three tests for more than one cluster is attached to
the one whose velocity it matches most closely.

\subsection{The weighting and the field}\label{app:weight} 
\textit{eROSITA} does not detect every young
star, so each attached star is used as a proxy for similar stars that go undetected. The
 scaling is set by the cluster members themselves: for a cluster with $N$ strict
members, of which $A$ are detected in X-rays, every attached star contributes to
all statistics with weight $N/A$. This weight is then scaled down by a field
purity factor, since some attached stars are actually field contaminants. The
purity is determined in regions without clusters: applying the same selection and attachment
 in parallax shells in front of and behind the complex yields the field
contribution per unit parallax, which is then interpolated to the cluster shell with
a power-law model. 

\subsection{The minimum membership and one cluster} A cluster with too few members cannot provide a reliable basis for attaching additional stars: when only a few members are available, individual stars determine its reach
and weights. Requiring at least 20 strict members eliminates seven clusters:
Pipe-North (12 strict members, where six stars at $17\degr$ separation
yielded $\rfifty = 132$\,pc), SO-5 (5), B30-splinter (11), $\lambda$~Ori
South-C (13), $\eta$~Cha (18), B59 (16), and Centaurus-Far (9). TWA-a meets
the threshold, but at 76\,pc its reach spans tens of degrees, exceeding the
linear perspective correction of Appendix~\ref{app:sigma}, and its corrected
dispersion (4.2\,\kms) is therefore not a meaningful measurement: the cluster is retained in $\rfifty$
and $\Pi$ but omitted from all dispersion and $\avir$ results.

\begin{figure}[t]
\centering
\includegraphics[width=\columnwidth]{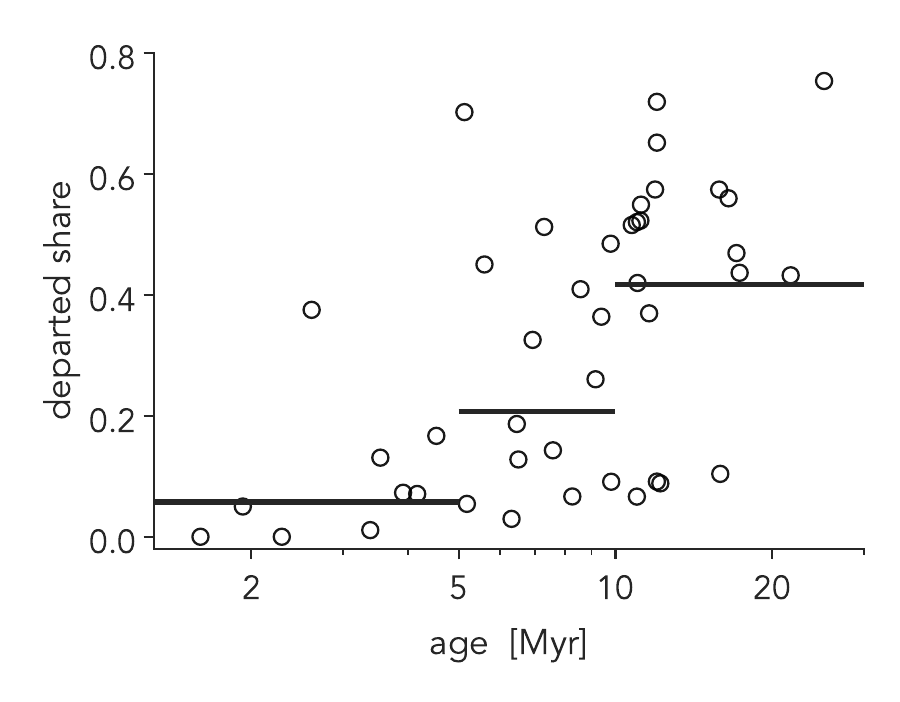}
\caption{The fraction of each Orion cluster's stars that have left it (the departed), against the cluster's age. Each circle is one cluster with at least five X-ray-detected members.  The black lines give the fraction in each of three age bins. Within a bin the counts from all clusters are added first and one fraction is taken from the totals, so the lines weight clusters by their number of stars rather than averaging the individual points in the figure.
}
\label{fig:fraction}
\end{figure}

\subsection{The halo} Figure~\ref{fig:fraction} shows the field-subtracted
share of the young X-ray population around the Orion clusters that the
algorithm does not assign: 0.06 at 1--5\,Myr, 0.21 at 5--10\,Myr and 0.42
at 10--30\,Myr. The share grows with age exactly as attrition predicts: the
longer a cluster has expanded, the larger the fraction of its stars beyond
the density contrast. The departed stars keep the birth dispersion,
$0.81^{+0.05}_{-0.05}$\,\kms\ at 5--10\,Myr and $0.93^{+0.03}_{-0.02}$ at
10--30\,Myr, while the assigned members of the same bins fall from 0.85
through 0.68 to 0.56 (the youngest bin holds 17 departed stars and constrains
little), so the assigned members are the slow subset of a population whose
fast tail is intact. They move outward: 82--93\% of them recede from their
cluster's center, against 70--73\% of the members, at a median outward speed
of 1.2--1.4\,\kms. Of the attached stars, 93\% in Orion and 95\% in
Sco--Cen pass every quality cut of the \texttt{SigMA} input, so they entered
the clustering. A density-contrast membership cannot contain stars that have
left the overdensity. In Sco--Cen the clusters subtend several degrees and
already hold their comoving population: the unassigned share is 0.05--0.09 at
every age, the extension changes little, and the Sco--Cen halo beyond that is
field the control removes.

\begin{figure*}[t]
\centering
\includegraphics[width=0.98\textwidth]{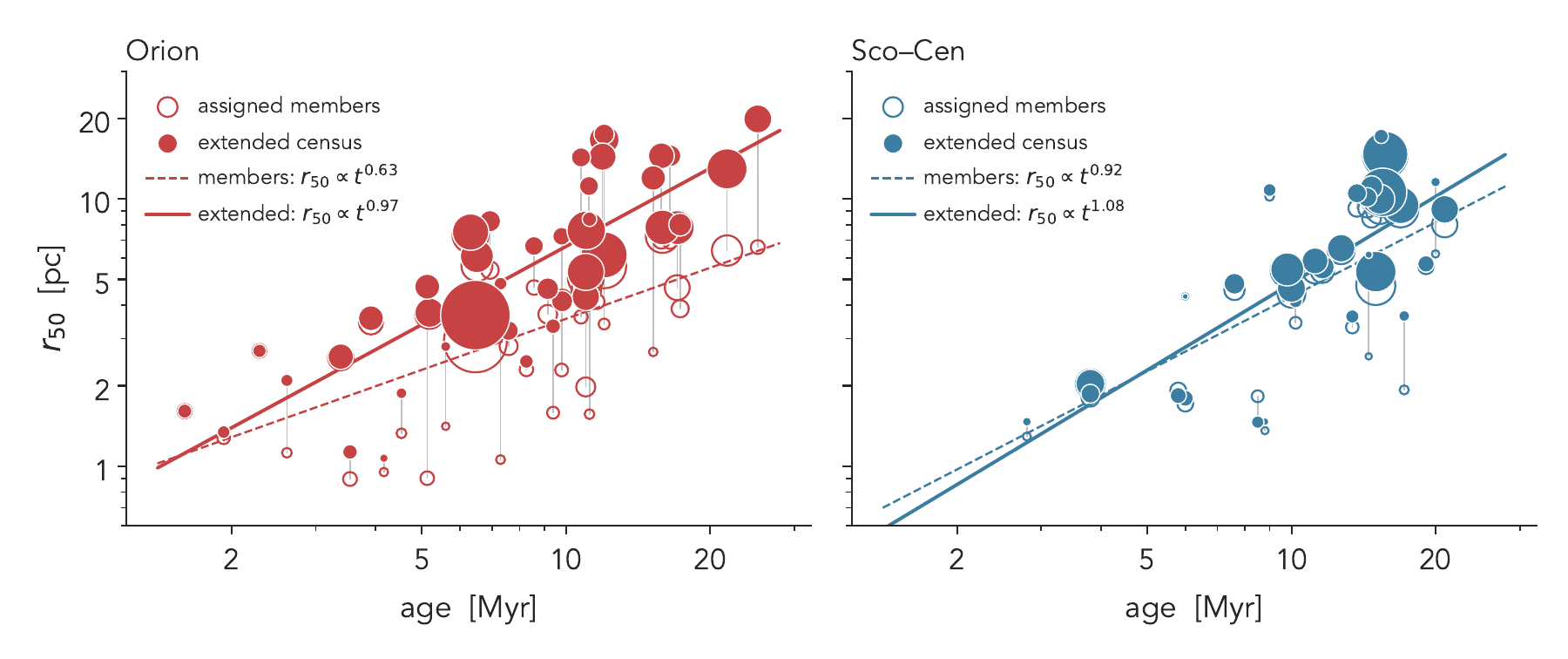}
\caption{Each cluster on its assigned members (open) and on the extended
census (filled) in the $\rfifty$--age plane, joined by a vertical line. The
dashed and solid lines are the members-only and the extended-census fits.}
\label{fig:arrows}
\end{figure*}

\subsection{What the members alone give} Figure~\ref{fig:arrows} shows every
cluster before and after the extension. On assigned members the same 75
clusters give $p = 0.63$ in Orion and 0.92 in Sco--Cen, a richness coupling
$b = 0.32$--0.37, and fixed-richness dispersion slopes of
$-0.24^{+0.05}_{-0.05}$ and $-0.26^{+0.08}_{-0.08}$. The members-only coefficient of $\sigbirth t$ is 0.43 in Orion and 0.63 in Sco--Cen, so a size-based kinematic age, the radius divided by the expansion rate, returns about half the isochronal age on members alone, which is why older clusters have been found kinematically younger than their isochrones \citep{luhman_census_2023, armstrong_expansion_2024, armstrong_expansion_2026}. The extension moves the
old clusters up in radius, the exponents to $0.97$ and $1.08$, the coupling to
0.13--0.14, and the dispersion slopes to zero (Sect.~\ref{sec:sigma}).

\subsection{Robustness} The extension rests on three choices: (1) the
factor of three in the ballistic reach, (2) the 3\,\kms\ velocity window,
and (3) the definition of $\sigbirth$. Each is varied here. The growth
exponents do not depend on any of them. The coefficient of $\sigbirth t$
depends on all three, in the first two cases because they set how much of a
cluster's outer population is admitted, and in the third by definition.

(1) The reach. With the factor at two, three and four, the Orion growth
exponent is 0.84, 0.97 and 0.94, the Sco--Cen one 1.02, 1.08 and 1.05, and
the departed share of the oldest bin moves only from 0.37 to 0.43. The
exponent holds because the reach grows in proportion to the age: a narrower
reach clips the same outer fraction of a 3-Myr cluster as of a 20-Myr one,
so it lowers all radii by one factor and leaves the slope alone. That factor
is the coefficient, which rises from 0.63 through 0.75 to 0.87 in Orion as
the reach widens, because each wider reach admits more of the outermost
stars. It has not converged at four, so the Letter's 0.75 is a lower limit
and is quoted with the window that produced it.

(2) The velocity window. Widening it from 3 to 5\,\kms\ admits the faster
departed stars, which lie farther out. The exponents move by at most 0.03,
to 1.00 and 1.09, the coefficient rises to 0.80, and the departed share of
the oldest bin to 0.46.

(3) The birth dispersion. The coefficient is $\rfifty/(\sigbirth t)$, so it
scales inversely with whatever $\sigbirth$ is taken to be, while the
exponents do not depend on it at all. The Letter defines $\sigbirth$ on the
assigned members of the clusters younger than 5\,Myr, the population
closest to birth. Defining it instead as the median dispersion of the whole
extended census gives 0.76 and 0.66\,\kms\ and moves the coefficient to
0.86 and 0.72. In Sco--Cen the young-member value rests on three clusters,
$\rho$~Oph (0.88\,\kms), Cha-1 (0.67) and Cha-2 (0.48), which span nearly a
factor of two, whereas the Orion value rests on nine.

\subsection{The surface-density exponent} The surface density inside the
half-member radius, $\sigfifty = (N/2)/(\pi\rfifty^{2})$, is the observable
of surface-density censuses, and the extended census gives
$\sigfifty \propto t^{-\gamma}$ with $\gamma = 1.40^{+0.20}_{-0.18}$ for
the two complexes fitted with one exponent and a free normalization each
($1.32^{+0.21}_{-0.19}$ in Orion alone, $1.59^{+0.44}_{-0.43}$ in Sco--Cen),
against $1.08^{+0.19}_{-0.17}$ on assigned members. Because $\sigfifty$ is
built from $N$ and $\rfifty$, the three slopes are related by arithmetic
rather than independently measured: $\gamma = 2p - d\log N/d\log t$. The
extended census has $d\log N/d\log t = +0.62^{+0.13}_{-0.13}$ in Orion and
$+0.56^{+0.39}_{-0.38}$ in Sco--Cen, a property of which clusters remain
detectable at each age rather than of any cluster's own history
(Appendix~\ref{app:limit}), so $\gamma$ sits below the free-expansion value
of 2 by exactly this richness trend.

\section{The \texttt{SigMA} detection boundary}
\label{app:limit}
\label{appendix:SigMA-boundary}

As a mode-seeking algorithm, \texttt{SigMA} only detects clusters that are denser than their surroundings.
A mode, meaning a local density maximum, is recognized by the algorithm as a cluster candidate only when the ratio of its density to that at the saddle point (highest minimum density) separating it from a neighboring mode is significant at a level $\alpha$ in the five-dimensional phase space of three positional and two tangential motion coordinates, scaled to a common metric \citep{ratzenbock_significance_2023, rottensteiner_orion_2026}. The coordinates differ between the two applications: Cartesian positions and tangential velocities are used for Sco--Cen, whereas spherical coordinates, parallax, and proper motions are used for clustering in Orion. Inverting the density significance test yields the minimum resolvable contrast of the algorithm as
\begin{align}\label{eq:density-contrast} 
\rho_{\text{mode}} / \rho_{\text{saddle}} >  \exp(z_{1-\alpha} / (p\sqrt{k/2})),
\end{align}
where $p=5$ is the number of dimensions and $k$ denotes the number of nearest neighbors (smoothing factor). For the \texttt{SigMA} run settings for Orion ($\alpha = 0.01$, $k \in [15,20,25,30]$) this contrast corresponds to 1.13--1.19 ($\sim$13--19\%) \citep{rottensteiner_orion_2026}. The clusters recovered in practice are not, however, concentrated against that limit. Taking the ratio between the highest and lowest member density as a lower bound on the mode-to-saddle contrast, the 46 published Orion clusters with at least ten members span 1.4--10.2. Their median is 3.1, and 89\% are more than 1.5 times above the formal threshold. Even the least-contrasted B30-splinter with 11 members exceeds it by 20\%. Contrast correlates with richness (Spearman $\rho=+0.65$, $p<10^{-4}$), so the sparsest clusters are indeed the ones sitting closest to the limit. Repeating the clustering with $k=10$--$25$ changes the distribution of mode contrasts by less than 10\%, smaller than its own spread. Thus, the significance test does not determine inclusion in the catalog. The limit is likely set further down the pipeline by the requirement that a cluster reproduce across resamplings and survive noise removal.

Because the operative limit is not the significance criterion, it cannot be derived from the algorithm, only measured. 
Eq.~(\ref{eq:density-contrast}) describes a ratio, so the same cluster is recoverable in a sparse field and lost in a crowded one, and no single volume density describes the boundary. However, we can still characterize where each census stops in practice for the datasets used here. Distances come from parallaxes, and parallax errors stretch every cluster along the line of sight. The median per-star distance error is 10.8~pc in the Orion shell against 0.96~pc in the nearer Sco--Cen. Thus, the depth the algorithm sees is
\begin{align}
L=\sqrt{(2k_{\rm proj}r_{50})^2+\sigma_d^2},\qquad  n=\frac{2\Sigma_{50}}{L},
\end{align}
 with $k_{\rm proj}=1.305$ converting a projected radius to a three-dimensional half-member radius for a Plummer profile (Sect.~\ref{sec:bound}) and $\Sigma_{50}$ the surface density of members within $r_{50}$. The least dense clusters recovered have $n=0.0030$ and $0.0034$ members pc$^{-3}$ in Orion and Sco--Cen, respectively, across a factor 2.8 in distance.
 Since $\sigma_d$ is already absorbed into $L$, that agreement is a consistency check on the correction rather than independent evidence about the algorithm, and $n$ should be read as an empirical completeness floor for the clustering solutions used here, not as a threshold. Solving at fixed $N$ converts it to 0.044--0.092 pc$^{-2}$ for clusters of 30 to 300 members. This boundary explains the one demographic trend in the census: an expanding cluster's density falls with age, so it drifts toward the boundary, and a poor cluster reaches it before a rich one, consistent with the contrast--richness correlation above. At older ages, the census keeps only its richer clusters, and the detected richness rises with age ($\mathrm{d}\log N/\mathrm{d}\log t=+0.6$, Appendix~\ref{app:member}) without any cluster gaining a single member. If the cluster population itself terminated near this density, the detected richness would be independent of age. The observed dependence, therefore, indicates a detection limit rather than a physical lower bound.

The age baseline needed to tell a power law from an exponential comes from
outside the two complexes. The corrected Akaike criterion (AICc) scores each
fitted form by its likelihood, with a penalty for extra parameters: a
difference of about 10 is decisive, and a difference of 1 to 2 separates
nothing. Over the single decade of age that Orion and Sco--Cen span, the two
forms tie, $\Delta$AICc = 1.6 in favor of the power law. The Orion run,
however, also recovers two older populations in the same volume, the Snake
\citep{tian_discovery_2020} and the BBJ filament \citep{beccari_uncovering_2020},
nine clusters each at 23.5--54\,Myr. With their assigned members extending the baseline to 54\,Myr, the
power law wins decisively: $\Delta$AICc = 10.8 over the four cohorts
together, and 10.4 over the three of the Orion volume. The reason is the
shape of the law itself. A power law $t^{-\gamma}$ declines at the
instantaneous fractional rate $\gamma/t$, so an exponential fitted to any
stretch of it returns an e-folding time near that sample's mean age divided
by $\gamma$: Orion gives $8.0^{+1.6}_{-1.3}$\,Myr against a mean age of
9.6\,Myr. A fitted timescale that follows the age of whatever sample it is
given is a property of the window rather than of the clusters.

\end{appendix}

\end{document}